\documentclass[a4paper,conference]{IEEEtran} 
\usepackage{epsfig,graphicx}
\usepackage{ setspace}
\usepackage{hyperref}
\usepackage[noadjust]{cite}
\usepackage{txfonts}

\newcommand{\kav}{{\bar{k}}}

\newcommand{\vsp}{\vspace*{3mm}}
\newcommand{\be}{\begin{equation}}
\newcommand{\ee}{\end{equation}}
\newcommand{\bd}{\begin{displaymath}}
\newcommand{\ed}{\end{displaymath}}

\newcommand{\bra}{\langle}
\newcommand{\ket}{\rangle}

\newcommand{\bc}{\mbox{\boldmath $c$}}

\newcommand{\bk}{\mbox{\boldmath $k$}}

\newcommand{\bA}{\mbox{\boldmath $A$}}

\newcommand{\rme}{{\rm e}}
\newcommand{\here}{\makebox(0,0)}
\newcommand{\bcirc}{\circle*{11}}

\begin{document}

\title{Controlled Markovian dynamics of graphs: \\unbiased generation of random graphs with 
prescribed topological properties}

\author{\authorblockN{ES Roberts}
\authorblockA{
Randall Division of Cell and \\
Molecular Biophysics\\
King's College London\\
Email: 
\href{mailto:ekaterina.roberts@googlemail.com}
{ekaterina.roberts@kcl.ac.uk}
}
\and
\authorblockN{A Annibale}
\authorblockA{
Dept. of Mathematics \\
King's College London\\
\href{mailto:alessia.annibale@kcl.ac.uk}
{alessia.annibale@kcl.ac.uk}
}
\and
\authorblockN{ACC Coolen}
\authorblockA{
Dept. of Mathematics and\\
Randall Division of Cell and \\
Molecular Biophysics\\
King's College London\\
\href{mailto:ton.coolen@kcl.ac.uk}
{ton.coolen@kcl.ac.uk}
}
}

\maketitle

\begin{abstract}
We analyze the properties of degree-preserving Markov chains based on elementary edge switchings in undirected and directed graphs. 
We give exact yet simple formulas for the 
mobility of a graph (the number of possible moves) in terms of its adjacency matrix. 
This formula allows us to define acceptance probabilities for edge switchings, such that the Markov chains become controlled Glauber-type 
detailed balance processes, designed to evolve to any required invariant measure 
(representing the asymptotic frequencies with which the allowed graphs are visited during the process).
\end{abstract}

\IEEEpeerreviewmaketitle

\section{Introduction}
\label{sec:Introduction}
Sampling uniformly the space of graphs with prescribed macroscopic properties has become a prominent 
problem in many application areas where tailored graph ensembles are used as proxies or null models for 
real complex networks. 
Quantities measured in real networks are often compared with the values these quantities take in their 
randomised counterparts. Ensembles of randomised networks allow one
to put error bars on these values, and therefore to identify which topological features of 
real networks deviate significantly from the null model. Such features are likely to reflect e.g. design principles or
evolutionary history of the network.
Ensembles of randomised graphs are  to be generated numerically, and each graph realisation should 
be produced with a probability proportional to a prescribed statistical weight (often taken to be uniform) of the graph, for the analysis to be unbiased.
At the level where the constraints of the randomized graphs involve only the simplest quantities, i.e. the degree sequence, even uniform generation 
of such graphs is known to be a non-trivial problem \cite{BenCan78,MolRee95,ChuLu02,StaBar05}. 
One classical algorithm for generating random networks with prescribed degrees 
\cite{NewStrWat01} assigns to each node a number of `edge stubs' equal to 
its desired degree, and joins iteratively pairs of randomly picked stubs to form a link. A drawback of this algorithm 
is the need for rejection of forbidden graphs (those with multiple edges or self-loops), which can lead to biased sampling  
\cite{KleHar11}. 
A second popular method for generating random graphs with a given degree sequence is `edge swapping', 
which involves successions of ergodic graph randomising moves that leave all degrees invariant \cite{EggHol81,Taylor81}. However, naive accept-all edge swapping 
will again cause sampling biases. The reason is that 
the {\em number} of edge swaps that 
can be executed is not a constant, it depends on the graph $\bc$ at hand; graphs which allow for many moves will  be generated more often. Any bias in the sampling of graphs invalidates their use as null models, so one is forced to mistrust all papers in which observations in real graphs have been tested against null models generated either via the 
`stubs' method or via randomisation by `accept-all edge swapping'.  
This situation can only be remedied via a systematic study of stochastic Markovian graph dynamics, which is the topic of this paper. We determine the appropriate adjustment to the probability of accepting a randomly chosen proposed edge swap for the process 
to visit each graph configuration with the same probability. This can be done for nondirected and directed graphs. 
We will also show that our method can be used to generate graphs with prescribed degree correlations.

\section{Generating random graphs via Markov chains}

A general and exact method for generating graphs from the set $G[\bk]=\{\bc\in G|~\bk(\bc)=\bk\}$ randomly (where $\bk$ denotes the degree sequence, or the joint in- and out-degree sequence of the graph), with specified 
probabilities $p(\bc)=Z^{-1}\exp[-H(\bc)]$ was developed in \cite{CooDemAnn09}. 
It has the form of a Markov chain:
 \begin{eqnarray}
\forall\bc\in G[\bk]:~~~~ p_{t+1}(\bc)&\!=\!\!\!&\sum_{\bc^\prime\in G[\bk]} W(\bc|\bc^\prime)p_t(\bc^\prime)
 \label{eq:MarkovChain}
 \end{eqnarray}
Here $p_t(\bc)$ is the probability of observing graph $\bc$ at time $t$ in the process,  
 and $W(\bc|\bc^\prime)$ is the one-step transition probability from graph $\bc^\prime$ to $\bc$.
For any set $\Phi$ of {\it ergodic} reversible elementary moves $F:G[\bk]\to G[\bk]$ we can choose transition probabilities of the form  
 \begin{eqnarray}
W(\bc|\bc^\prime)&\!\!=\!\!& \sum_{F\in\Phi} q(F|\bc^\prime)\Big[\delta_{\bc,F\bc^\prime} A(F\bc^\prime|\bc^\prime)+\delta_{\bc,\bc^\prime} [1\!-\!A(F\bc^\prime|\bc^\prime)]\Big]
\nonumber\\
\label{eq:TransitionProbabilities}
\end{eqnarray}
The interpretation is as follows. 
At each step a candidate move $F\in\Phi$ is drawn with probability $q(F|\bc^\prime)$, where $\bc^\prime$ denotes the current graph. 
This move is accepted (and the move $\bc^\prime\to\bc=F\bc^\prime$ executed) with probability $A(F\bc^\prime|\bc^\prime)\in[0,1]$, 
which depends on the current graph $\bc^\prime$ and the proposed new graph $F\bc^\prime$. If the move is rejected, 
which happens with probability $1\!-\!A(F\bc^\prime|\bc^\prime)$, the system stays in $\bc^\prime$. 
We may always exclude from $\Phi$ the identity operation. 
One can prove that the process  (\ref{eq:MarkovChain}) will converge towards the equilibrium measure $p_\infty(\bc)=Z^{-1}\exp[-H(\bc)]$ 
upon making in (\ref{eq:TransitionProbabilities}) 
the choices
\begin{eqnarray}
q(F|\bc)&=& I_F(\bc)/n(\bc)
\label{eq:candidates}
\\[-0.5mm]
A(\bc|\bc^\prime)&=&
\frac{n(\bc^\prime)\rme^{-\frac{1}{2}[H(\bc)-H(\bc^\prime)]}}{
n(\bc^\prime)e^{-\frac{1}{2}[H(\bc)-H(\bc^\prime)]}+
n(\bc)\rme^{\frac{1}{2}[H(\bc)-H(\bc^\prime)]}}
\label{eq:acceptance}
\end{eqnarray}
Here $I_F(\bc)=1$ if the move $F$ can act on graph $\bc$, $I_F(\bc)=0$ otherwise,  
and $n(\bc)$ denotes the total number of moves that can act on a graph $\bc$ (the `mobility' of state $\bc$):
\begin{equation}
n(\bc)=\sum_{F\in\Phi} I_F(\bc).
\end{equation}

\section{Degree-constrained dynamics of nondirected graphs}

We  first apply our results to algorithms that randomise undirected graphs, while conserving all degrees, by 
application of edge swaps that act on quadruplets of nodes and their mutual links. Such moves were shown to be ergodic, i.e. any two graphs with the same 
degree sequence can be connected by a finite number ofsuccessive edge swaps \cite{Taylor81,EggHol81}.  

Let us define 
the set $Q=\{(i,j,k,\ell)\in\{1,\ldots,N\}^4|~i\!<\!j\!<\!k\!<\!\ell\}$ of all ordered node quadruplets. The possible edge 
swaps to act on $(i,j,k,\ell)$ are the following, with thick lines indicating existing links and thin lines indicating absent links that will be 
swapped with the existing ones, 
and where (IV, V, VI) are the inverses of (I, II, III):
\vspace*{10mm}

\setlength{\unitlength}{0.10mm}
\hspace*{15mm}
\begin{picture}(500,100)(5,0)
\put(50,175){\here{I}}
\put(5,0){\here{\bcirc}}\put(105,0){\here{\bcirc}}\put(5,100){\here{\bcirc}}\put(105,100){\here{\bcirc}}
\put(0,125){\here{$i$}}\put(100,125){\here{$j$}}\put(100,-25){\here{$k$}}\put(0,-25){\here{$\ell$}}
\thinlines \put(0,0){\line(0,1){100}} \put(100,0){\line(0,1){100}}
\thicklines\put(0,100){\line(1,0){100}}\put(0,101){\line(1,0){100}}\put(0,99){\line(1,0){100}}
\put(0,0){\line(1,0){100}} \put(0,-1){\line(1,0){100}}\put(0,1){\line(1,0){100}}
\put(250,175){\here{II}}
\put(205,0){\here{\bcirc}}\put(305,0){\here{\bcirc}}\put(205,100){\here{\bcirc}}\put(305,100){\here{\bcirc}}
\put(200,125){\here{$i$}}\put(300,125){\here{$j$}}\put(300,-25){\here{$k$}}\put(200,-25){\here{$\ell$}}
\thinlines \put(200,0){\line(1,1){100}} \put(300,0){\line(-1,1){100}}
\thicklines \put(200,100){\line(1,0){100}} \put(200,101){\line(1,0){100}} \put(200,99){\line(1,0){100}}
\put(200,0){\line(1,0){100}}\put(200,1){\line(1,0){100}}\put(200,-1){\line(1,0){100}}
\put(450,175){\here{III}}
\put(405,0){\here{\bcirc}}\put(505,0){\here{\bcirc}}\put(405,100){\here{\bcirc}}\put(505,100){\here{\bcirc}}
\put(400,125){\here{$i$}}\put(500,125){\here{$j$}}\put(500,-25){\here{$k$}}\put(400,-25){\here{$\ell$}}
\thinlines \put(400,0){\line(0,1){100}} \put(500,0){\line(0,1){100}}
\thicklines \put(400,100){\line(1,-1){100}}\put(400,101.5){\line(1,-1){100}}\put(400,98.5){\line(1,-1){100}}
\put(400,0){\line(1,1){100}}\put(400,1.5){\line(1,1){100}}\put(400,-1.5){\line(1,1){100}}
\end{picture}
\\[1.6cm]
\begin{picture}(500,100)(-200,0)
\put(50,175){\here{IV}}
\put(5,0){\here{\bcirc}}\put(105,0){\here{\bcirc}}\put(5,100){\here{\bcirc}}\put(105,100){\here{\bcirc}}
\put(0,125){\here{$i$}}\put(100,125){\here{$j$}}\put(100,-25){\here{$k$}}\put(0,-25){\here{$\ell$}}
\thicklines \put(0,0){\line(0,1){100}} \put(100,0){\line(0,1){100}}
\put(1,0){\line(0,1){100}} \put(101,0){\line(0,1){100}}
\put(-1,1){\line(0,1){100}} \put(99,0){\line(0,1){100}}
\thinlines\put(0,100){\line(1,0){100}} \put(0,0){\line(1,0){100}}
\put(250,175){\here{V}}
\put(205,0){\here{\bcirc}}\put(305,0){\here{\bcirc}}\put(205,100){\here{\bcirc}}\put(305,100){\here{\bcirc}}
\put(200,125){\here{$i$}}\put(300,125){\here{$j$}}\put(300,-25){\here{$k$}}\put(200,-25){\here{$\ell$}}
\thicklines \put(200,0){\line(1,1){100}} \put(300,0){\line(-1,1){100}}
\put(201,-1){\line(1,1){100}} \put(299,-1){\line(-1,1){100}}
\put(199,1){\line(1,1){100}} \put(301,1){\line(-1,1){100}}
\thinlines \put(200,100){\line(1,0){100}}
\put(200,0){\line(1,0){100}}
\put(450,175){\here{VI}}
\put(405,0){\here{\bcirc}}\put(505,0){\here{\bcirc}}\put(405,100){\here{\bcirc}}\put(505,100){\here{\bcirc}}
\put(400,125){\here{$i$}}\put(500,125){\here{$j$}}\put(500,-25){\here{$k$}}\put(400,-25){\here{$\ell$}}
\thicklines \put(400,0){\line(0,1){100}} \put(500,0){\line(0,1){100}}
\put(401,0){\line(0,1){100}} \put(501,0){\line(0,1){100}}
\put(399,0){\line(0,1){100}} \put(499,0){\line(0,1){100}}
\thinlines \put(400,100){\line(1,-1){100}}
\put(400,0){\line(1,1){100}}
\end{picture}
\\

We group the edge swaps into three pairs (I,IV), (II,V), and (III,VI), and label all three resulting auto-invertible operations for each ordered quadruple $(i,j,k,\ell)$
with a subscript $\alpha$. Our auto-invertible edge swaps are now written as $F_{ijk\ell;\alpha}$, with  $i<j<k<\ell$ and $\alpha\in\{1,2,3\}$. We define 
associated indicator
functions $I_{ijk\ell;\alpha}(\bc)\in\{0,1\}$ that detect whether (1) or not (0) the edge swap $F_{ijk\ell;\alpha}$ can act on state $\bc$, so
\begin{eqnarray}
\hspace*{-5mm}I_{ijk\ell;1}(\bc)= c_{ij}c_{k\ell}(1\!-\!c_{i\ell})(1\!-\!c_{kj})+(1\!-\!c_{ij})(1\!-\!c_{k\ell})c_{i\ell}c_{kj}
\label{eq:Fcond1}
\\[-0.5mm]
I_{ijk\ell;2}(\bc)= c_{ij}c_{\ell k}(1\!-\!c_{ik})(1\!-\!c_{\ell j})+(1\!-\!c_{ij})(1\!-\!c_{\ell k})c_{ik}c_{\ell j}
\label{eq:Fcond2}
\\[-0.5mm]
I_{ijk\ell;3}(\bc)= c_{ik}c_{j\ell}(1\!-\!c_{i\ell})(1\!-\!c_{jk})+(1\!-\!c_{ik})(1\!-\!c_{j\ell})c_{i\ell}c_{jk}
\label{eq:Fcond3}
\end{eqnarray}
If 
$I_{ijk\ell;\alpha}(\bc)=1$, this edge swap will operate as follows:
\begin{eqnarray}
F_{ijk\ell;\alpha}(\bc)_{qr}&\!\!= 1\!-\!c_{qr}~~~& {\rm for}~(q,r)\in {\cal S}_{ijk\ell;\alpha}
\label{eq:F1}
\\[-0.5mm]
F_{ijk\ell;\alpha}(\bc)_{qr}&\!\!= c_{qr}~~~& {\rm for}~(q,r)\notin {\cal S}_{ijk\ell;\alpha}
\label{eq:F2}
\end{eqnarray}
where
\begin{eqnarray}
{\cal S}_{ijk\ell;1}&=& \{(i,j),(k,\ell),(i,\ell),(k,j)\}
\label{eq:S1}
\\
{\cal S}_{ijk\ell;2}&=& \{(i,j),(\ell,k),(i,k),(\ell,j)\}
\label{eq:S2}
\\
{\cal S}_{ijk\ell;3}&=& \{(i,k),(j,\ell),(i,\ell),(j,k)\}
\label{eq:S3}
\end{eqnarray}
Insertion of these definitions into the recipe (\ref{eq:TransitionProbabilities},\ref{eq:candidates},\ref{eq:acceptance})
gives
\begin{eqnarray}
W(\bc|\bc^\prime)&\!=\!&\! \sum_{i<j<k<\ell}~\sum_{\alpha\leq 3} \frac{I_{ijk\ell;\alpha}(\bc^\prime)}{n(\bc^\prime)}
\label{eq:edgeswaptransitions}
\\
&&\hspace*{-1cm}\times
\Big[
\frac{\delta_{\bc,F_{ijk\ell;\alpha}\bc^\prime} \rme^{-\frac{1}{2}[E(F_{ijk\ell;\alpha}\bc^\prime)-E(\bc^\prime)]}
+\delta_{\bc,\bc^\prime} \rme^{\frac{1}{2}[E(F_{ijk\ell;\alpha}\bc^\prime)-E(\bc^\prime)]}}{
\rme^{-\frac{1}{2}[E(F_{ijk\ell;\alpha}\bc^\prime)-E(\bc^\prime)]}+
\rme^{\frac{1}{2}[E(F_{ijk\ell;\alpha}\bc^\prime)-E(\bc^\prime)]}}
\Big]\nonumber
\end{eqnarray}
with $E(\bc)=H(\bc)+\log n(\bc)$. 
The process (\ref{eq:edgeswaptransitions}) can be described as the following algorithm. Given an instantaneous
graph $\bc^\prime$: (i) pick uniformly at random a quadruplet $(i,j,k,\ell)$ of sites, (ii) if at least one of the three edge swaps $\bc^\prime\to \bc= F_{ijk\ell;\alpha}(\bc^\prime)$ is possible, 
select  one of these uniformly at random and 
execute it with an acceptance probability
\begin{eqnarray}
A(\bc|\bc^\prime)&=& \big[1+\rme^{E(F_{ijk\ell;\alpha}\bc^\prime)-E(\bc^\prime)}\big]^{-1}
\label{eq:canonical_acceptance}
\end{eqnarray}
then return to (i). 
For this Markov chain recipe to be practical we finally need a formula for the mobility $n(\bc)$ of a graph.
This could be calculated \cite{CooDemAnn09}, giving\footnote{Efficient expressions for the change in mobility following one move have also been derived \cite{CooDemAnn09}, to avoid unnecessary matrix multiplication in the computational implementation. 
}
(with ${\rm Tr}\bA=\sum_i A_{ii}$):
\begin{eqnarray}
n(\bc)
&=&
\frac{1}{4}\big(\!\sum_i k_i\big)^2 +\frac{1}{4}\!\sum_{i}k_i
-\frac{1}{2}\!\sum_{i}k_i^2
-\frac{1}{2}\!\sum_{ij}k_{i}c_{ij}k_j
\nonumber\\
&&+\frac{1}{4}{\rm Tr}(\bc^4)+\frac{1}{2}{\rm Tr}(\bc^3)~~~~
\label{eq:nc}
\end{eqnarray}
Naive `accept-all' edge swapping, where $A(\bc|\bc')=1$, corresponds to choosing $E(\bc)=0$ in (\ref{eq:edgeswaptransitions}), and 
  would give the {\em biased} graph sampling probabilities $p_\infty(\bc)=n(\bc)/\sum_{\bc^\prime}n(\bc^\prime)$ upon equilibration.
The graph mobility acts as an entropic force which can only be  neglected if (\ref{eq:nc})  is dominated by its first three terms; it was shown that 
a sufficient condition for this to be true is 
$\bra k^2\ket k_{\rm max}/\bra k\ket^2 \!\ll\! N$. 
In networks with narrow degree sequences this condition holds, and naive edge swapping is roughly acceptable. 
However, one has to be careful with scale-free graphs, where $\bra k^2\ket$ and  $k_{\rm max}$ diverge as $N\to\infty$.

\section{Degree-constrained dynamics of directed graphs}
For directed networks 
the generalisation of the canonical edge swap (up to relabelling of nodes) is shown below:
\vspace*{5mm}

{\setlength{\unitlength}{0.10mm}
\hspace*{16mm}
\begin{picture}(500,140)(5,-20)
\put(205,0){\here{\bcirc}}\put(305,0){\here{\bcirc}}\put(205,100){\here{\bcirc}}\put(305,100){\here{\bcirc}}
\put(200,125){\here{$x_i$}}\put(300,125){\here{$x_j$}}\put(300,-25){\here{$y_j$}}\put(200,-25){\here{$y_i$}}
\thinlines \put(200,0){\vector(1,1){100}} \put(200,101){\vector(1,-1){100}}
\thicklines \put(200,100){\vector(1,0){100}} \put(200,101){\vector(1,0){100}} \put(200,99){\vector(1,0){100}}
\put(200,0){\vector(1,0){100}}\put(200,1){\vector(1,0){100}}\put(200,-1){\vector(1,0){100}}
\end{picture}
\\[5mm]$~~$}
%
\noindent
This move samples the space of all directed graphs with 
prescribed in- and out- degrees ergodically only if self-interactions are permitted \cite{Rao96}.
We could now proceed with our formalism as before, but it is more efficient in the case of directed graphs to define the swaps and the indicator function in terms of pairs of links rather than quadruples of nodes. 

Let $\bc$ now be a nonsymmetric connectivity matrix, such that 
$c_{ab}=1$ if $a\rightarrow b$, otherwise $c_{ab}=0$. Consider $\Lambda$ to be the set of links within the network defined by $\bc$. We write $\mathbf{x} = (x_i, x_j)\in \Lambda$ if and only if  $c_{x_i x_j}=1$.
The indicator function can be written as 
$$I_{\mathbf{x} , \mathbf{y};\square} = \left\{
\begin{array}{l l c }   
    1& \mbox{ if } \mathbf{x} , \mathbf{y} \in \Lambda \mbox{ and }(x_i, y_j),  (y_i, x_j)  \notin \Lambda \\
    0& \mbox{ otherwise }
\end{array}\right.$$
If $I_{\mathbf{x} , \mathbf{y};\square} = 1$, then the corresponding autoinvertible operation $F_{\mathbf{x} , \mathbf{y};\square}$ acts on state $\bc$  as follows:
\begin{eqnarray}
F_{\mathbf{x} , \mathbf{y};\square}(\bc)_{qr} = 1-c_{qr}~
 &{\rm for}~q\in \{x_i, y_i \}~ {\rm and}~r\in \{x_j, y_j \}
\label{eq:square1}
\\[-0.5mm]
F_{\mathbf{x} , \mathbf{y};\square}(\bc)_{qr}= c_{qr}~~~ &{\rm otherwise}
\label{eq:square2}
\end{eqnarray}
When self-interactions are forbidden, a further type of elementary move is required to ensure ergodicity \cite{Rao96}. It can be visualised as reversing a
triangle cycle:

\hspace*{5mm}\setlength{\unitlength}{0.10mm}
\begin{picture}(200,200)(-100,-50)
\put(0,0){\here{\bcirc}}\put(100,0){\here{\bcirc}}\put(100,0){\here{\bcirc}}
\put(-25,-15){\here{$x_i$}}\put(150,-5){\here{$x_j=y_i$}}\put(15,115){\here{$y_j$}}
\thicklines\put(0,0){\vector(1,0){100}}\put(100,0){\vector(-1,1){100}}\put(-5,100){\vector(0,-1){100}}
\put(0,-1){\vector(1,0){100}}\put(101,1){\vector(-1,1){100}}\put(-4,100){\vector(0,-1){100}}
\put(0,1){\vector(1,0){100}}\put(99,-1){\vector(-1,1){100}}\put(-6,100){\vector(0,-1){100}}
\put(300,0){\here{\bcirc}}\put(400,0){\here{\bcirc}}\put(300,100){\here{\bcirc}}
\put(275,-15){\here{$x_i$}}\put(450,-5){\here{$x_j=y_i$}}\put(315,115){\here{$y_j$}}
\thicklines\put(295,0){\vector(0,1){100}}\put(400,0){\vector(-1,0){100}}\put(300,100){\vector(1,-1){100}}
\put(296,1){\vector(0,1){100}}\put(400,1){\vector(-1,0){100}}\put(301,101){\vector(1,-1){100}}
\put(294,-1){\vector(0,1){100}}\put(400,-1){\vector(-1,0){100}}\put(299,99){\vector(1,-1){100}}
\end{picture}

\noindent
The indicator function for this new move is
$$I_{\mathbf{x} , \mathbf{y};\triangle} = \left\{
\begin{array}{l l c }   
    1& \mbox{if } \mathbf{x} , \mathbf{y},(y_j, x_i) \in \Lambda ~\rm{and}~ x_j=y_i\\
      &  \mathbf{x}^{-1} , \mathbf{y}^{-1}, (x_i,y_j)  \notin \Lambda \\
    0& \mbox{ otherwise }
\end{array}\right.$$
where $\mathbf{x}^{-1}=(x_j,x_i)$ represents a link that is to be reversed. 
The corresponding autoinvertible operation represented by $F_{\mathbf{x} , \mathbf{y};\triangle} $ acts on state $\bc$ as follows:
\begin{eqnarray}
F_{\mathbf{x} , \mathbf{y};\triangle} (\bc)_{qr}&= 1-c_{qr}~~~& {\rm for}~(q,r)\in {\cal S}_{x_i, x_j, y_j}
\label{eq:triangle1}
\\[-0.5mm]
F_{\mathbf{x} , \mathbf{y};\triangle} (\bc)_{qr}&\!\!= c_{qr}~~~& {\rm for}~(q,r)\notin {\cal S}_{x_i, x_j, y_j}
\label{eq:triangle2}
\end{eqnarray}
in which ${\cal S}_{a b c}= \{(a,b),(b,c),(c,a), (b,a),(c,b),(a,c)\}$ is the set of pairs of vertices of the triangle. 

We now combine edge swaps and triangular reversions into (\ref{eq:TransitionProbabilities}). This requires us to express $ \sum_{F\in\Phi} q(F|\bc^\prime)$ from the point of view of sampling links $\mathbf{x}$ and $\mathbf{y}$. This is in fact straightforward, as each pair of links can either trigger the square indicator function (if they define an edge swap), a triangle indicator function (if they define a triangle reversion), or neither (when the indicator function returns zero). It only remains to observe the symmetries of the problem: cycling through each pair of bonds we will come across each unique edge swap twice, and across each unique triangle swap three times
\cite{RobCoo11}. Hence  
$$  \sum_{F\in\Phi} q(F|\bc^\prime)= \sum_{\mathbf{x}, \mathbf{y} \in \Lambda} \frac{\frac{1}{2}I_{\mathbf{x} , \mathbf{y};\square} + \frac{1}{3}I_{\mathbf{x} , \mathbf{y};\triangle} }{n(\bc^\prime)} $$ 
which expresses sampling over all possible moves in terms of (the more tangible) sampling over links. 
With the above definitions in hand, the transition probability is given by
\cite{RobCoo11}
\begin{eqnarray}
W(\bc|\bc^\prime)&=& \sum_{\mathbf{x}, \mathbf{y} \in \Lambda} \frac{\frac{1}{2}I_{\mathbf{x} , \mathbf{y};\square} + \frac{1}{3}I_{\mathbf{x} , \mathbf{y};\triangle} }{n(\bc^\prime)}
\label{eq:directedW}
\\
&&\hspace*{-5mm}\times
\Big[
\frac{\delta_{\bc,F_{\mathbf{x}, \mathbf{y}}\bc^\prime} \rme^{-\frac{1}{2}[E(F_{\mathbf{x}, \mathbf{y}}\bc^\prime)-E(\bc^\prime)]}
+\delta_{\bc,\bc^\prime} \rme^{\frac{1}{2}[E(F_{\mathbf{x}, \mathbf{y}}\bc^\prime)-E(\bc^\prime)]}}{
\rme^{-\frac{1}{2}[E(F_{\mathbf{x}, \mathbf{y}}\bc^\prime)-E(\bc^\prime)]}+
\rme^{\frac{1}{2}[E(F_{\mathbf{x}, \mathbf{y}}\bc^\prime)-E(\bc^\prime)]}}
\Big]
\nonumber
\end{eqnarray}
where $n(\bc)= n_\square(\bc)+n_\triangle(\bc)$ is the total 
number of valid moves that can act on the directed graph $\bc$, written as the sum of the number of edge swaps and the number of triangle reversions.  
The mobility terms can again be calculated, giving \cite{RobCoo11}
\begin{eqnarray*}
n_\square(\bc) &=& \frac{1}{2}   Tr(\bc \bc^T \bc \bc^T)
 - \! \!  \sum_{i,j} \mathbf{k}^{out}_i c_{ij} \mathbf{k}^{in}_j 
+  Tr(\bc \bc^T \bc)
\nonumber\\ 
&&+ \frac{1}{2} (\sum_i \mathbf{k}_i)^2
+ \frac{1}{2}  Tr(\bc^2)
 - \! \! \sum_i \mathbf{k}^{out}_j \mathbf{k}^{in}_j 
\\
n_\triangle(\bc)
&=&  \frac{1}{3} 
	Tr(\mathbf{\bc}^3) 
	- Tr(\mathbf{c^{\updownarrow}} \mathbf{c}^2 ) 
	+ Tr(\mathbf{c^{\updownarrow}}^2 \mathbf{c})
	- \frac{1}{3} Tr(\mathbf{c^{\updownarrow}}^3)
\end{eqnarray*}
with $(\bc^T)_{ij}=c_{ji}$ and 
 $(\bc^\updownarrow)_{ij}=c_{ij}c_{ji}$. Also for directed graphs 
it is possible to calculate efficient expressions for how the mobilities change following a single move. 

For the simplest case where $H(\bc)$ is constant, i.e. the Markov chain is to evolve towards a flat measure, we immediately observe that the required move acceptance probabilities are
 \begin{eqnarray}
A(\bc|\bc^\prime)&=& \big[1+\frac{n_\square(\bc) + n_\triangle(\bc)}{n_\square(\bc^\prime) + n_\triangle(\bc^\prime)}\big]^{-1}
\label{eq:canonical_acceptance_simple}
\end{eqnarray}
\vsp

\section{Numerical examples for directed graphs}

Examples illustrating the canonical Markov chains for nondirected graphs are given in \cite{CooDemAnn09}. Here we compare two variants of the {\em directed} edge rewiring algorithm:
the naive `accept all moves' version, and the version with the canonical mobility 
corrections  (\ref{eq:canonical_acceptance_simple}).
Consider a network with $K+2$ nodes, of which one has degrees $(k^{\rm in},k^{\rm out})=(0,K)$, 
one has $(k^{\rm in},k^{\rm out})=(K,0)$, and the remaining $K$ have degrees $(k^{\rm in},k^{\rm out})=(1,1)$.
Self-interactions are forbidden.
Up to relabelling, there are two such graph types (see below):\vspace*{6mm}

\begin{minipage}[b]{0.445\linewidth}
\hspace*{-5mm}\includegraphics[width = \linewidth]{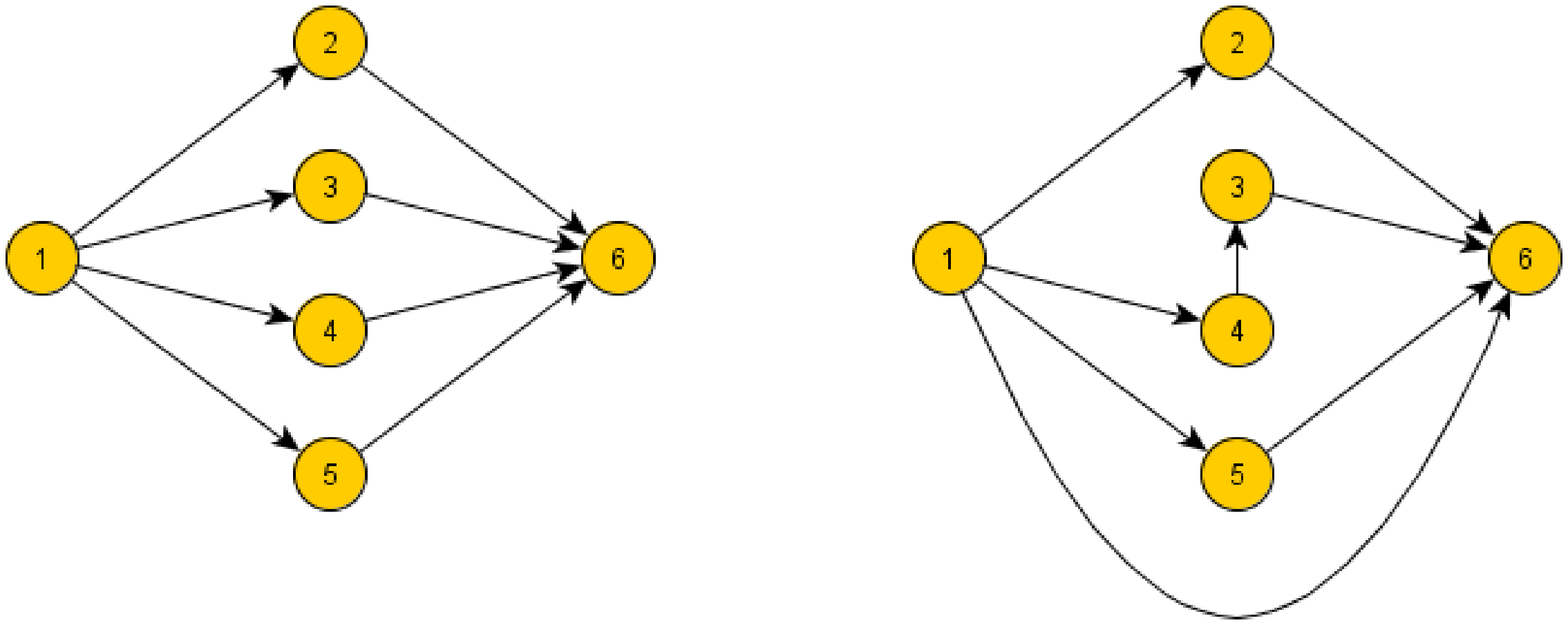} 
\end{minipage}
\hspace*{-5mm}
\begin{minipage}[b]{0.44\linewidth}
\begin{tabular}{ | c | c | c |}
\hline 
 & accept all	& correct \\
K=25& process	& process \\
& avg. $n(\bc)$	&avg. $n(\bc)$ \\
 \hline 
predict	& 58.52	& 47.92 \\
actual 	& 58.32	& 47.95 \\
\hline 
\end{tabular}
\vspace{2pt}
\end{minipage}
\vspace*{4mm}

\noindent
The left graph type has mobility $K(K-1)$, and only occurs in one network configuration. 
The right graph type has mobility of $2K-3$, and multiplicity $K(K-1)$. 
The table shows the results of numerical simulations for $K=25$, compared to theoretical predictions. 
We have used the mobility itself as a marker for the proportion of time spent in each type of configuration; the difference between the two processes is striking, and will translate into differences in measurements of any observable which differentiates between the two configurations. 

\section{Generation of random graphs with prescribed degree correlations via rewiring algorithms}
\vspace*{-2mm}
Our approach can be extended to accurately target desired degree correlations defined by
$W(k,k')$.
This can be achieved for undirected graphs by ensuring convergence of the above Markov chain to the following non-uniform measures 
\cite{PerCoo08},
\begin{eqnarray}
&&\hspace*{-8mm}
p(\bc)\!=\!\frac{\delta_{\!\bk,\bk(\bc)}}{Z}\prod_{i<j}\left[\frac{\bra k\ket
}{N}\frac{W(k_i,k_j)}{p(k_i)p(k_j)}\delta_{c_{ij},1}+\Big(1\!-\!\frac{\bra k\ket
}{N}\frac{W(k_i,k_j)}{p(k_i)p(k_j)}\Big)\delta_{c_{ij},0}\right]
\nonumber\\[-1.5mm]
&&\label{eq:newer_connectivity}
\end{eqnarray}
with $p(k)=N^{-1}\sum_i \delta_{k_i,k}$, $W(k,k^\prime)=(N\bra k\ket)^{-1}\sum_{ij}c_{ij}\delta_{k_i,k}\delta_{k_j,k^\prime}$, 
 $\bra k\ket=\sum_k kp(k)$,  and
\be
W(k)=\sum_{k'}W(k,k')=p(k)k/\bra k\ket
\label{eq:marginalW}
\ee
In the language of our process, (\ref{eq:newer_connectivity}) 
corresponds to the choice
\begin{eqnarray}
&&\hspace*{-0.7cm}H(\bc)\!=\!-\sum_{i<j}\log \Big[\frac{\bra k\ket
}{N}\frac{W(k_i,k_j)}{p(k_i)p(k_j)}\delta_{c_{ij},1}+\Big(1\!-\!\frac{\bra k\ket
}{N}\frac{W(k_i,k_j)}{p(k_i)p(k_j)}\Big)\delta_{c_{ij},0}\Big]
\nonumber
\end{eqnarray}
Hence, for the candidate edge swaps $\bc'\to \bc=F_{ijk\ell;\alpha}\bc'$ 
the acceptance probability (\ref{eq:acceptance}) can be used, where
\begin{eqnarray}
e^{H(\bc)-H(\bc')}&=& 
\prod_{(a,b)\in S_{ijk\ell;\alpha}}
\Big[L_{ab}\delta_{c_{ab},1}+L^{-1}_{ab}\delta_{c_{ab},0}\Big]
\label{eq:PC_deltaH}
\end{eqnarray}
with
$
L_{ab}= N\kav/[\Pi(k_a,k_b)k_a k_b]-1
$
and relative degree correlations $\Pi(k,k')=W(k,k')/W(k)W(k')$.

\begin{figure}[t]
\vspace*{-1mm} \hspace*{8mm} 
\setlength{\unitlength}{0.41mm}
\begin{picture}(200,210)
\put(-10,120){\includegraphics[height=75\unitlength]{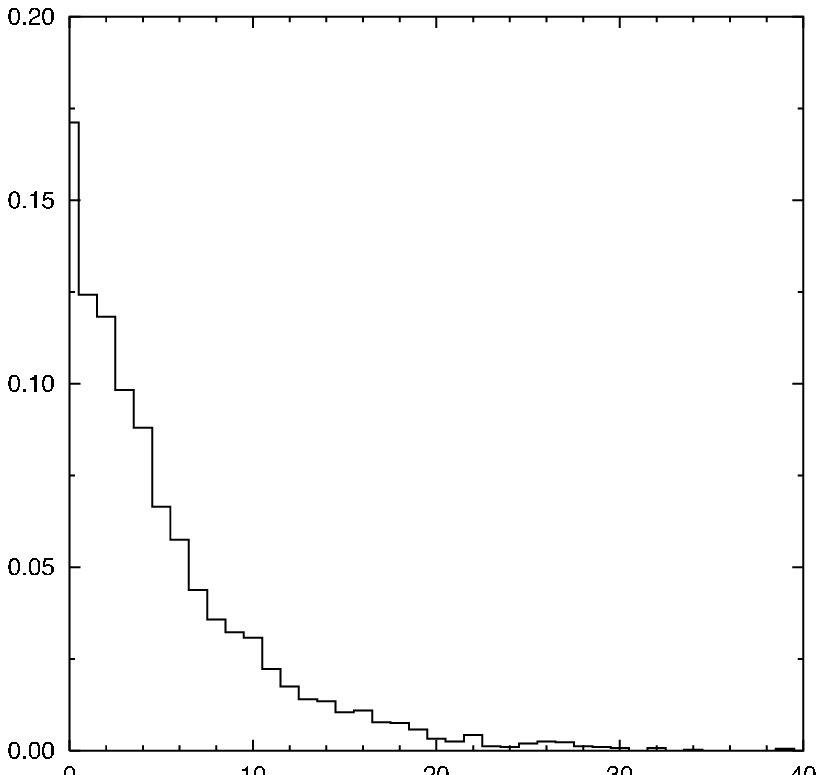}}
\put(-20,163){\footnotesize $P(k)$}
\put(30,110){\footnotesize $k$}
\put(100,123){\includegraphics[height=75\unitlength]{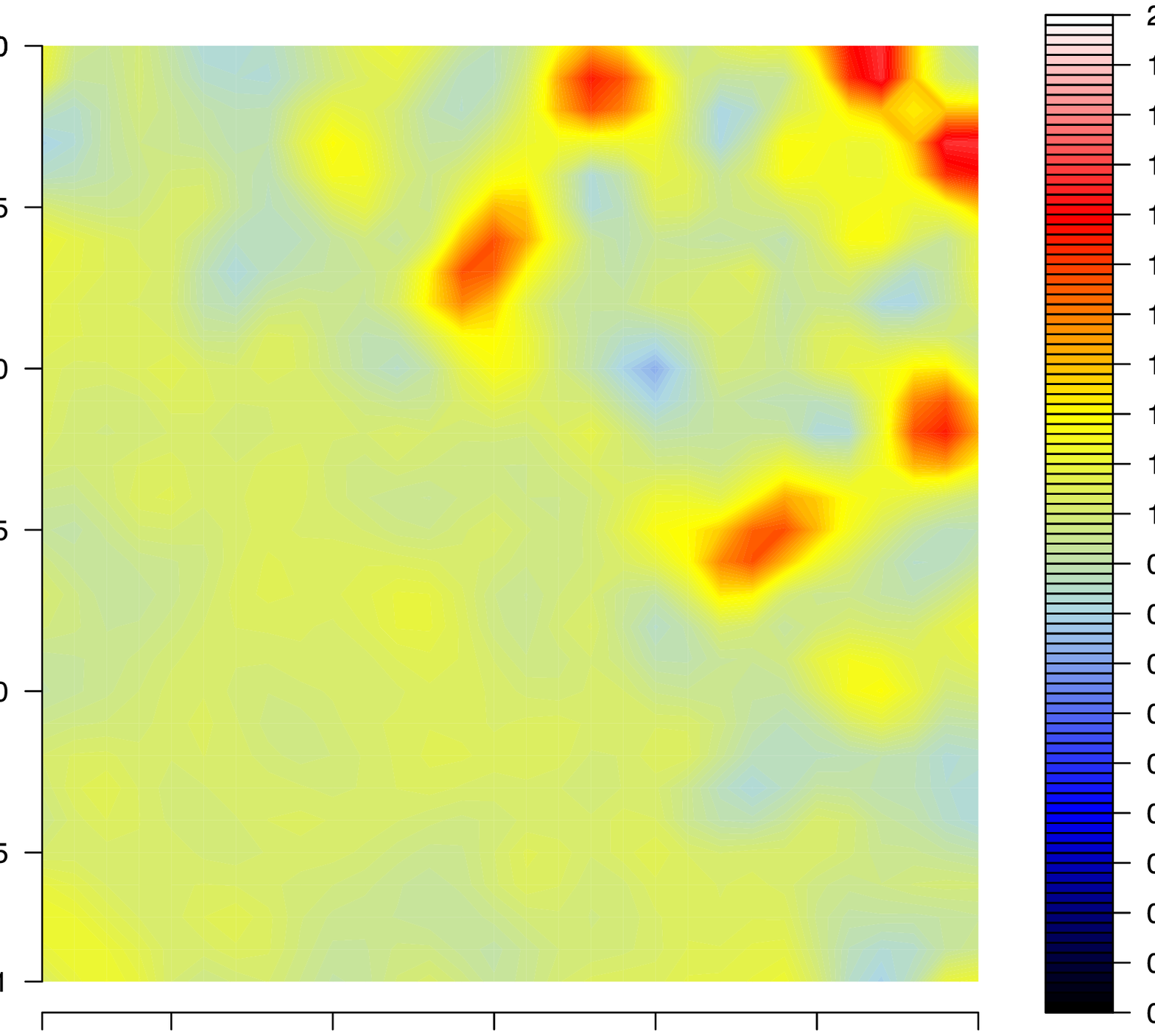}}
\put(90,163){\footnotesize $k^\prime$}
\put(140,110){\footnotesize $k$}
\put(120,200){\small $\Pi(k,k^\prime|\bc_0)$}
\put(-5,15){\includegraphics[height=75\unitlength]{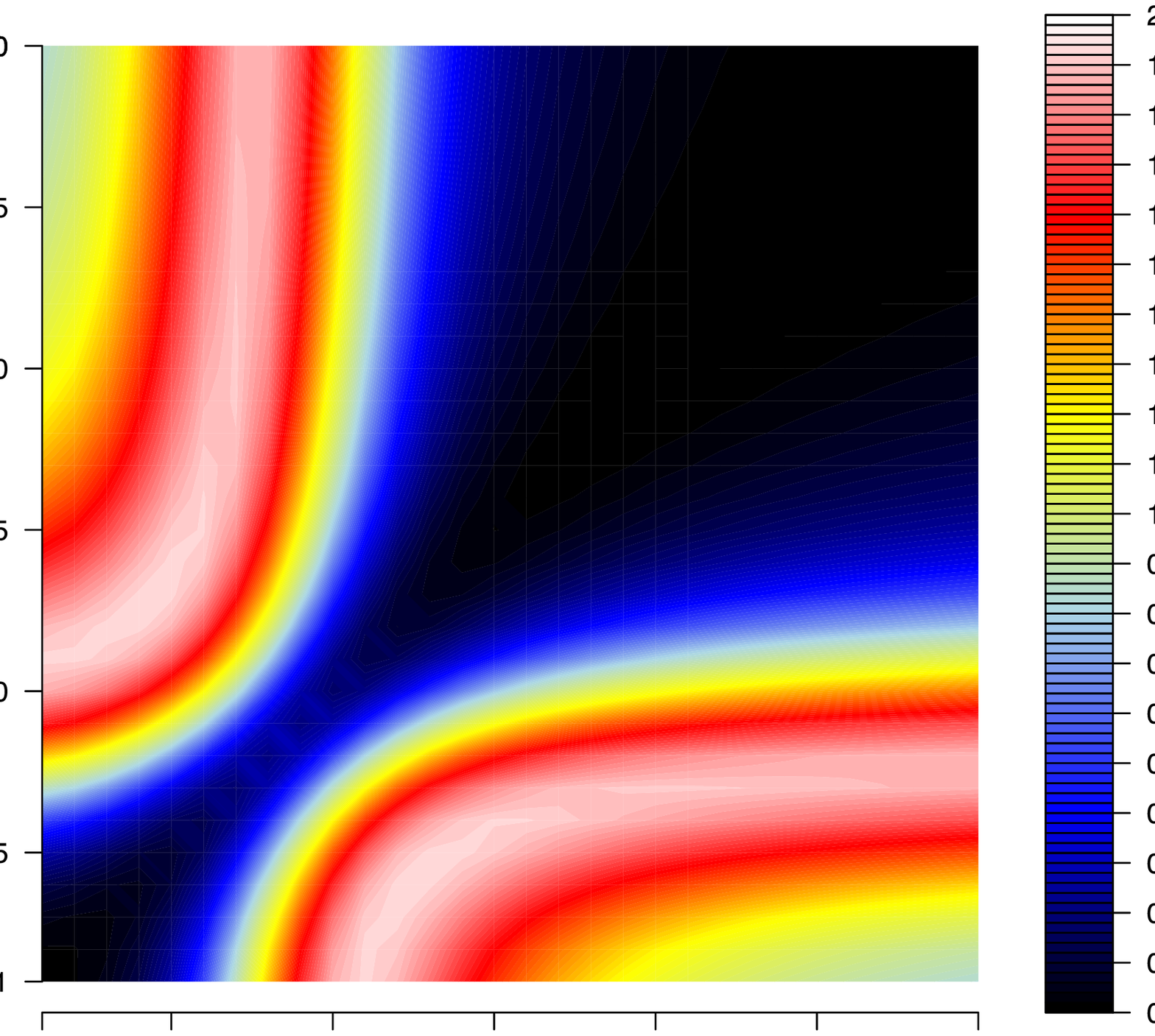}}
\put(-20,60){\footnotesize $k^\prime$}
\put(30,0){\footnotesize $k$}
\put(10,95){\small $\Pi(k,k^\prime)$ (theory)}
\put(100,15){\includegraphics[height=75\unitlength]{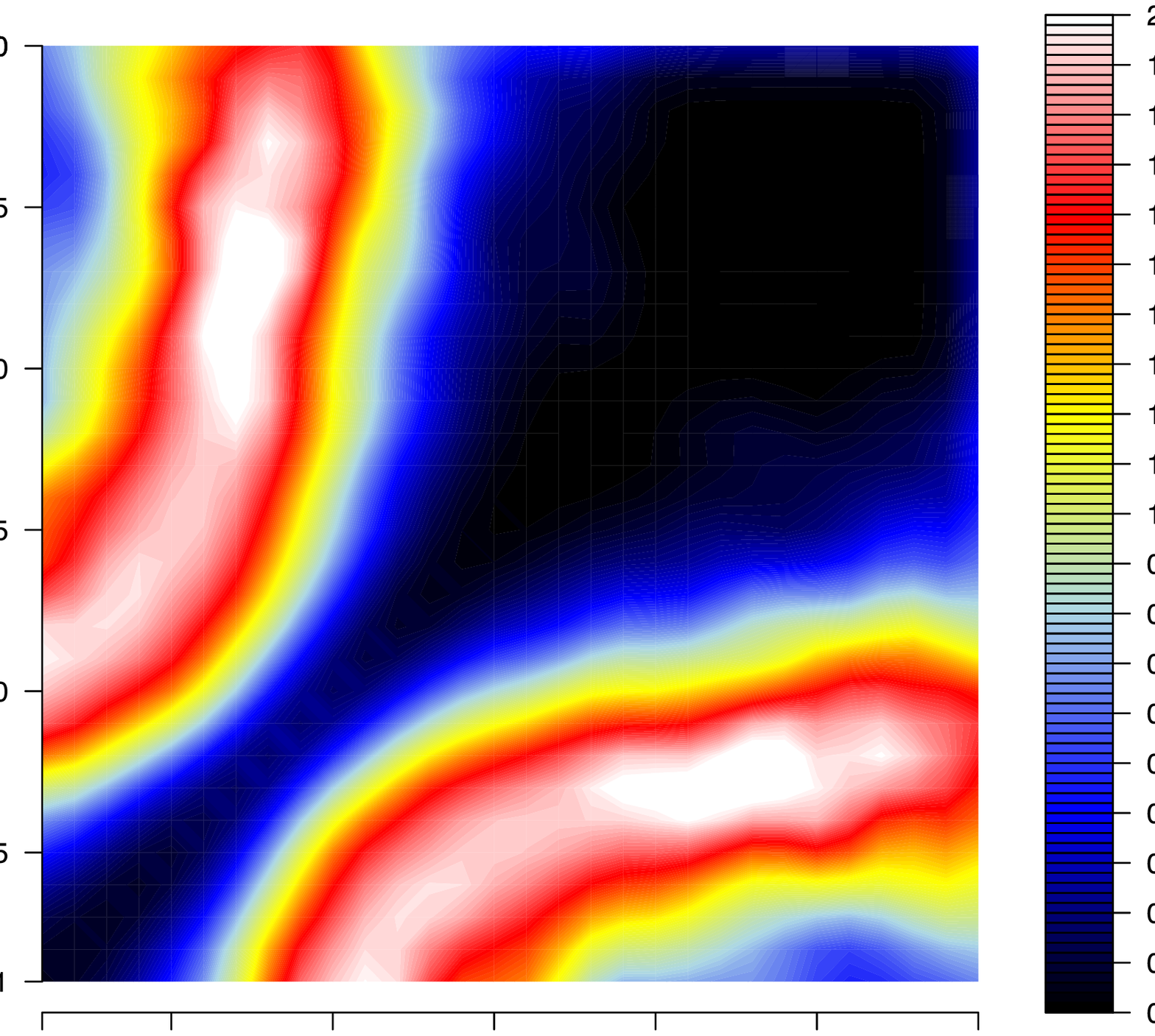}}
\put(90,60){\footnotesize $k^\prime$}
\put(140,0){\footnotesize $k$}
\put(120,95){\small $\Pi(k,k^\prime|\bc_{\rm final})$}
\end{picture}
 \vspace*{-5mm}
\caption{Results of Markovian dynamics tailored to target the non-uniform measure
 (\ref{eq:newer_connectivity}). Top left: degree distribution of the (randomly generated) initial graph $\bc_0$, with $N=4000$ and $\bra k\ket=5$.
Top right: $\Pi(k,k^\prime|\bc_0)$ of the initial graph.
Bottom left: the target relative degree correlations (\ref{eq:Pi_predicted}) chosen in (\ref{eq:newer_connectivity}).
Bottom right: colour plot of $\Pi(k,k^\prime|\bc_{\rm final})$ in the final graph $\bc_{\rm final}$, measured after 75,000 accepted moves of the
Markov chain (\ref{eq:edgeswaptransitions}).} 
\vspace*{-6mm}
\label{fig:Pi}
\end{figure}
We show an example of the Markov chain (\ref{eq:edgeswaptransitions}) targeting (\ref{eq:newer_connectivity})
where relative degree correlations are chosen as
\vspace*{-1mm}
\begin{eqnarray}
\hspace*{-6mm}\Pi(k,k^\prime)\!\!&=&\!\!(k-k^\prime)^2/[\beta_1-\beta_2 k+\beta_3 k^2][\beta_1-\beta_2k^\prime+\beta_3 k^{\prime 2}]
\label{eq:Pi_predicted}
\end{eqnarray}
(the parameters $\beta_i$ follow from (\ref{eq:marginalW})).
An initial graph $\bc_0$ was constructed with a non-Poissonian degree 
distribution and no degree correlations, i.e. 
$\Pi(k,k^\prime|\bc_0)\approx 1$.
After iterating the Markov chain until equilibrium 
(after $75,000$ accepted moves, and after reaching maximal
Hamming distance between initial and final configuration)
degree correlations are seen in very good agreement 
with their target values; (see Figure \ref{fig:Pi}). 
Extension to directed graphs is achieved by replacing $k$ with 
$\bk=(k^{\rm in},k^{\rm out})$,
allowing repetition of site indices in
(\ref{eq:PC_deltaH})
and bearing in mind that $\Pi(\bk_a,\bk_b)\neq \Pi(\bk_b,\bk_a)$.

\section{Discussion}

In this paper we focused on
how to generate numerically tailored random graphs with controlled macroscopic structural properties, to serve e.g. as null models in hypothesis testing. Bias in the generation of random graphs has the potential to invalidate all further statistical 
analysis performed on the generated networks and has been well documented in the literature; it is known to affect the `stubs' method and the `accept-all' edge swapping method. However, the lack so far of workable corrections or alternatives has meant that these issues have often been ignored. 
Our theory offers a practical and theoretically sound approach to uniformly generating random graphs from ensembles which 
share certain topological characteristics with a real network, and can 
therefore serve as a reliable tool for building unbiased null models.  

\bibliographystyle{IEEEtran}
\bibliography{IEEEabrv,ref}

\end{document}